\documentclass[twocolumn,pra,superscriptaddress,showpacs,aps,floatfix]{revtex4-2}
\usepackage{color}
\usepackage{amsmath}
\usepackage{amssymb}
\usepackage{txfonts}
\usepackage{graphicx}
\usepackage{epstopdf}
\usepackage[unicode]{hyperref}
\hypersetup{pdftitle={2D SOC BEC},
 pdfauthor={},
 pdfsubject={2D SOC BEC},
  colorlinks=true,
    linkcolor=red,
    citecolor=magenta,
    filecolor=black,
    urlcolor=blue}

\begin{document}

\title{Exploring the Origin of CP Violation in the Standard Model}

\author{Chilong Lin}
\email{lingo@mail.nmns.edu.tw}

\affiliation{National Museum of Natural Science, 1st, Guan Chien RD., Taichung, 40453 Taiwan, ROC}

\date{Version of \today. }

\begin{abstract}

In this article, we present a very general but not ultimate solution of CPV problem in the standard model.
Our study starts from a naturally Hermitian ${\bf M^2}\equiv M^q \cdot M^{q\dagger}$ rather than the previously assumed Hermitian $M^q$.
The only assumption employed here is that the real part and imaginary part of $\bf M^2$ can be, respectively, diagonalized by a common $\bf U^q$ matrix.
Such an assumption leads to a $\bf M^2$ pattern which depends on only five parameters and can be diagonalized analytically by a $\bf U^q$ matrix which depends on only two of the parameters.
Two of the derived mass eigenvalues are predicted degenerate if one of the parameters $\bf C ~(C')$ in up- (down-) quark sector is zero.
As the $\bf U^q$ patterns are obtained, thirty-six $V_{CKM}$ candidates are yielded and only eight of them,
classified into two groups, fit empirical data within the order of $O(\lambda)$.
One of the groups is further excluded in a numerical test,
and the surviving group predicts that the degenerate pair in a quark type are the lightest and the heaviest generations rather than the lighter two generations assumed in previous researches.
However, there is still one unsatisfactory prediction in this research, a quadruple equality in which four CKM elements of very different values are predicted to be equal.
It indicates the $\bf M^2$ pattern studied here is still oversimplified by that employed assumption and the ultimate solution can only be obtained by diagonalizing the unsimplified  $\bf M^2$ matrix containing nine parameters directly.
The $V_{CKM}$ presented here is already very close to such an ultimate CPV solution.

\end{abstract}
\maketitle

%%%%%%%%%%%%%%%%%%%%%%%%%
%%%%  sec: intro   %%%%%%
%%%%%%%%%%%%%%%%%%%%%%%%%

\section{Introduction}

The theoretical origin of CP violation (CPV) is a long unsolved problem in particle physics since its first discovery in the decays of neutral kaons \cite{Christenson1964}.
We know that, in the standard model (SM) of electroweak interactions, this can only be yielded "explicitly" by a complex phase in the Cabibbo-Kobayashi-Maskawa (CKM) matrix \cite{Cabibbo1963, KM1973} which is a product of the two unitary transformation matrices $U^u$ and $U^d$ which diagonalize the mass matrices $M^u$ and $M^d$ of up- and down-type quarks, respectively.
However, even if there are only three already known fermion generations for now, a $3 \times 3$ mass matrix with eighteen unknown parameters is obviously too complicated to be diagonalized analytically.
Thus, for decades, an analytical solution of CPV problem remains obscured. \\

Besides the "explicit" way to bring complex phases into the theory, physicists proposed another way to bring complex phases into the theory by employing an extra Higgs doublet \cite{TDLee1973} and expect that the phase difference between vacuum expectation values (VEVs) of two Higgs doublets will be nonzero.
It's usually referred to as the "spontaneous" way to break CP symmetry.  \\

In principle, a most general $3 \times 3~M^q$ matrix has nine complex elements,
 and each of them has two coefficients, one from the real part and one from the imaginary part.
Thus, there are at most eighteen parameters in total in such a matrix.
If we can diagonalize it analytically, there are always possibilities for inducing complex phases into the CKM matrix by choosing parameters suitably.
However, such a matrix is obviously too complicated to be diagonalized analytically.
Thus, researchers proposed many ways to simplify the $M^q$ pattern to a manageable level.
For instance, various ansatz like Fritzsch ansatz (FA) \cite{Fritzsch1978, Fritzsch1979}, Cheng-Sher ansatz (CSA) \cite{Cheng1987}, Du-Xing ansatz (DXA) \cite{DuXing1993},
combination of the Fritzsch and the Du-Xing ansatz (DFXA and FDXA), combination of different assignments in the Du-Xing ansatz (${\tilde X}$A ), non-mixing top quark ansatz (NTA) [and references therein] \cite{Carcamo2007}, and Fukuyama-Nishiura ansatz (FNA) [and references therein] \cite{Matsuda2000} were imposed with $ad~hoc$ zeros in $M^q$ to simplify the pattern.
Instead, one may employ symmetries like $S_N$ symmetry among fermion generations \cite{Derman1979, Lee1986a, Lee1986b, Lin1988, Lin2020} and many others to build correlations among  $M^q$ elements so as to simplify its pattern.
However, assumptions, constraints, symmetries, or $ad~hoc$ zero elements always reduce the generality of researches.
Here we would like to present a very general solution of the CPV problem in SM and see how close it is to the ultimate one. \\

In one of our previous studies \cite{Lin2020},
an $S_2$-symmetric model gave us several complex CKM matrices with a predicted Jarlskog invariant \cite{Jarlskog1985}, an estimate of CPV strength,
which is four orders stronger than that detected in current experiments.
In an even earlier article \cite{Lin1988}, we found that an $S_3$-symmetric model would not give any CP-violating, complex phases in its CKM matrix.
None of them fit experiments very well even though \cite{Lin2020} gave a concrete proof of explicitly violated CP symmetry in SM with several evidences.
Comparing these two researches,
we observed that the predicted CKM matrix regularly deviates more from experiments if the constraints are stronger.
Thus, we raise the following question: if there is a model which is completely $S_N$-nonsymmetric, will its CKM elements fit experiments better?
Another reason that pushes us toward such a nonsymmetric study is that no $S_N$ symmetries had been observed in our present universe.
That explains why those $S_N$-symmetric predictions don't fit experiments well. \\

Following such a concept, we will start the study from a most general $3 \times 3$ $M^q$ pattern containing eighteen parameters.
Instead of assuming a Hermitian $M^q$ to simplify it down to only nine parameters \cite{Lin2019, Lin2020},
we study here a naturally Hermitian ${\bf M^2}\equiv M^q \cdot M^{q\dagger}$ matrix since fewer assumptions make the theory more general.
The only assumption employed in this article is a common $\bf U^q$ matrix which diagonalizes the real part $\bf M^2_R$ and the imaginary part $\bf M^2_I$ of $\bf M^2$ respectively and simultaneously.
In \cite{Lin2020} we employed two assumptions, a Hermitian $M^q$ and a common $U^q$, to achieve a set of analytical solutions.
Here, only the assumption of a common $\bf U^q$ remains.
It makes the solution thus obtained more general than that given in \cite{Lin2019} and more close to the reality.
 \\

In Section II, we start from a general review regarding CPV, CKM matrix and fermion mass matrices.
Then, we analyze the problems researchers addressed in previous investigations when trying to diagonalize the mass matrices
and present a way to solve those problems with an assumption much weaker than previous ones.
Subsequently, with the only assumption that the real part ${\bf M^2}_R$ and imaginary part ${\bf M^2_I}$ of ${\bf M^2}$ can be diagonalized simultaneously and respectively by the same ${\bf U^q}$ matrix, the ${\bf M^2}$ matrix is simplified down to a manageable level so as to be diagonalized analytically.
During the derivation, an interesting relation between ${\bf M^2_R}$ and ${\bf M^2_I}$ is employed by the assumption to bring about extra correlations among their elements
and thus reduce the parameter number in an ${\bf M^2}$ from nine down to five.
That relation was originally proposed in \cite{Branco1985} for the Natural-Flavor-Conservation (NFC) in two-Higgs-doublet models (2HDMs).
The eigenvalues and eigenvectors thus obtained look completely the same as those given in \cite{Lin2019},
but they are of very different implications.
Thus, we will denote them by boldfaced letters like ${\bf M^2}$ and ${\bf U^q}$ to distinguish them from those derived from the original, un-squared matrix $M^q$ given in \cite{Lin2019} and even earlier articles \cite{Lin1988, Lee1990, Lin1994b}.  \\

As the ${\bf U^q}$ matrices are obtained, surely the CKM matrix $V_{CKM}$ is obtained.
In section III, thirty-six $V_{CKM}$ candidates are presented since there are six ways to designate three mass eigenvalues to three physical quark masses in a quark type.
All of them are dependent on only four parameters since the eigenvectors of a quark type are dependent on only two of the five parameters.
Such a parameterization of $V_{CKM}$ is very natural since all its elements are directly dependent on Yukawa couplings and the VEV of its only Higgs doublet.
As to be shown latter, there are some predicted equalities among CKM elements, and they exclude some of the $V_{CKM}$ candidates.
Only eight of them, classified into two groups, fit experiments within the order of $O(\lambda)$, and only one of those two groups gives acceptable solutions in a numerical test.
In such a case, the derived mass spectrum predicts a degeneracy between the heaviest and lightest generations, say the up and top quarks in the up-quark sector,
when one of the parameters $\bf C$ is 0. \\

Though this research gives a better solution to the CPV problem than previous similar researches,
it's still not the ultimate one since there is still one assumption remains.
As mentioned above, there are several predicted equalities among CKM elements in such a model.
One of them is a quadruple equality which correlates four of the CKM elements, and this is obviously incoincident with empirical values.
This suggests that the employed assumption of a common $\bf U^q$ simplifies the pattern of $\bf M^2$ too much.
The problem will be completely solved if we can diagonalize the unsimplified $\bf M^2$ containing nine parameters directly.
Unfortunately, such an ultimate solution looks still far beyond our capability for now,
 and more discussions on this topic will be provided in section IV as conclusions. \\

%%%%%%%%%%%%%%%%%%%%%
%%%  subsec: HTM  %%%
%%%%%%%%%%%%%%%%%%%%%
\section{The Pattern of Mass Matrix}

In standard model, the only source of CPV is a complex phase in $V_{CKM}$, the CKM matrix.
Obviously, the most orthodox way to study the CPV problem is to find out how such a phase comes into $V_{CKM}$. \\

Starting from the Yukawa couplings of $Q$ quarks in SM which are usually given  by
\begin{equation}
-{\cal L}_Y ~=  \bar{Q_L} Y^{d} \Phi d_R +  \bar{Q_L} \epsilon Y^{u} \Phi^{\ast} u_R + h. c.,
 \end{equation}
where $Y^q$ are $3 \times 3$ Yukawa-coupling matrices for quark types $q=u$ and $d$, and $\epsilon$ is the $2 \times 2$ antisymmetric tensor.
$Q_L$ is left-handed quark doublets, and $d_R$ and $u_R$ are right-handed down- and up-type quark singlets in their weak eigenstates, respectively.  \\

When the Higgs doublet $\Phi$ acquires a vacuum expectation value,
$\langle \Phi \rangle = \left( \begin{array}{cc} 0 \\ v/\sqrt{2} \end{array}\right)$, Eq.(1) yields mass terms for quarks with $M^q=Y^q~ v / \sqrt{2}$ the mass matrices.
The physical states are obtained by diagonalizing $M^q$ with unitary matrices $U^q_{L,R}$, as $M^q_{diag.}=U^q_L \cdot M^q \cdot U^q_R = U^q_L \cdot (Y^q v/\sqrt{2} ) \cdot U^q_R $.
As a result, the charged-current $W^{\pm}$ interactions couple to the physical $u_L$ and $d_L$ quarks with couplings given by
\begin{equation}
-{\cal L}_{W^{\pm}} ~= {g \over \sqrt{2}} (\bar{u}_L,~\bar{c}_L,~\bar{t}_L)~\gamma^{\mu} W^+_{\mu}~ V_{CKM} \left( \begin{array}{ccc} d_L \\ s_L \\ b_L \end{array}\right) + h. c.,
 \end{equation}
 where
\begin{eqnarray}
V_{CKM} = U^u_L \cdot U^{d \dagger}_L =\left( \begin{array}{ccc} V_{ud} & V_{us} & V_{ub}  \\ V_{cd} & V_{cs} & V_{cb}  \\ V_{td} & V_{ts} & V_{tb} \end{array}\right).
\end{eqnarray}
Hereafter, the subindex $L$ in quark fields $q_L$ and unitary matrices $U_L$ will be neglected if not necessary. \\

As $V_{CKM}$ is a product of $U^u$ and $U^{d \dagger}$ which are derivatives of $M^u$ and $M^d$ matrices, respectively,
obviously the mass matrices decide everything in $V_{CKM}$ including its phases.
Thus, the most natural way to study the origin of CPV in SM shall start from the patterns of $M^u$ and $M^d$. \\

As mentioned above, a most general $3\times 3$ matrix contains eighteen parameters, and such a matrix is obviously too complicated to be diagonalized directly.
In several of our previous researches on $S_N$ symmetries \cite{Lin1988, Lee1986a, Lee1986b, Lee1990, Lin1994b},
mass matrices were naturally Hermitian as demanded by the $S_N$ invariance.
However, in several recent researches \cite{Lin2019, Lin2020}, $M^q =M^{q \dagger}$ is an assumption employed to simplify $M^q$ and consequently give us complex, CP-violating CKM matrices.
These two series of researches studied the CPV problem from very different aspects.
However, the $M^q$ patterns obtained in \cite{Lin2020} through a purely numeric derivation were revealed to possess $S_2$ symmetries between two of the three fermion generations.
They started from different ends of the problem and finally reached the same goal.
However, such an assumption is still uncomfortable to us.
We would like to further drop off that assumption and study this topic in a non-Hermitian basis for a better generality.   \\

In case fermions have only three generations, the general pattern of a $3 \times 3$ mass matrix is then given by
\begin{eqnarray}
M^q =  \left( \begin{array}{ccc} A_1 +i D_1 & B_1 +i C_1 & B_2+i C_2 \\  B_4 + i C_4 & A_2 +i D_2 & B_3 +i C_3 \\ B_5 +i C_5 & B_6 +i C_6 & A_3 +i D_3 \end{array}\right) ,
\end{eqnarray}
where all $A$, $B$, $C$, and $D$ parameters are by definition real, and there are eighteen of them in total. \\

Multiplying such a $M^q$ with $M^{q\dagger}$ will receive an $\bf M^2$ matrix given by
\begin{eqnarray}
{\bf M^2 =  M^2_R + M^2_I =} \left( \begin{array}{ccc} {\bf A_1} & {\bf B_1} & {\bf B_2} \\  {\bf B_1} & {\bf A_2} & {\bf B_3} \\ {\bf B_2}  & {\bf B_3}  & {\bf A_3}  \end{array}\right)
   + i \left( \begin{array}{ccc} {\bf 0} & {\bf C_1} & {\bf C_2}  \\  {\bf -C_1} & {\bf 0} & {\bf C_3}  \\ {\bf -C_2} & {\bf -C_3} & {\bf 0} \end{array}\right),
\end{eqnarray}
in which $\bf M^2_R$ and $\bf M^2_I$ are, respectively, the real part and imaginary part of $\bf M^2$, and $\bf M^2$ is naturally Hermitian.
The boldface parameters ${\bf A}$, ${\bf B}$, and ${\bf C}$ are defined by
\begin{eqnarray}
{\bf A_1} &=& A_1^2 + D_1^2 + B_1^2 + C_1^2 + B_2^2 + C_2^2, \\
{\bf A_2} &=& A_2^2 + D_2^2 + B_3^2 + C_3^2 + B_4^2 + C_4^2, \\
{\bf A_3} &=& A_3^2 + D_3^2 + B_5^2 + C_5^2 + B_6^2 + C_6^2, \\
{\bf B_1} &=& A_1 B_4 + D_1 C_4 + B_1 A_2 + C_1 D_2 + B_2 B_3 +C_2 C_3, \\
{\bf B_2} &=& A_1 B_5 + D_1 C_5 + B_1 B_6 + C_1 C_6 + B_2 A_3 +C_2 D_3, \\
{\bf B_3} &=& B_4 B_5 + C_4 C_5 + B_6 A_2 + C_6 D_2 + A_3 B_3 +D_3 C_3, \\
{\bf C_1} &=& D_1 B_4 -A_1 C_4 +A_2 C_1 -B_1 D_2 +B_3 C_2 -B_2 C_3, \\
{\bf C_2} &=& D_1 B_5 -A_1 C_5 +B_6 C_1 -B_1 C_6 +A_3 C_2 -B_2 D_3, \\
{\bf C_3} &=& C_4 B_5 -B_4 C_5 +D_2 B_6 -A_2 C_6 +A_3 C_3 -B_3 D_3.
\end{eqnarray}
In such a case, only nine parameters are independent in a $\bf M^2$ matrix. \\

Since the $U^q_L$ matrix of $M^q$ is the same as the $\bf U^q$ of ${\bf M^2}$,
surely it's much easier to diagonalize an ${\bf M^2}$ matrix with nine parameters than an $M^q$ matrix with eighteen parameters.
However, the ${\bf M^2}$ pattern in Eq.(5) is still too complicated to be diagonalized analytically.
Here we would like to employ an assumption that ${\bf M^2_R}$ and ${\bf M^2_I}$ can be diagonalized by the same ${\bf U^q}$ to build extra relations among parameters and thus simplify ${\bf M^2}$ down to a manageable level.
It is the only assumption employed in this research.
Obviously, it's more general than previous similar researches which employed more assumptions or $ad~hoc$ constraints.   \\

For two arbitrary matrices $M_1$ and $M_2$, if there is the same $U$ matrix which diagonalize them both simultaneously and respectively,
there exists an interesting relation:
\begin{eqnarray}
M_1 \cdot M_2^\dagger -  M_2 \cdot M_1^\dagger =0,
\end{eqnarray}
which was originally proposed in \cite{Branco1985} to solve the FCNC problem in 2HDMs by surveying matrix pairs which are diagonalized by the same $U$. \\

Letting $M_1 ={\bf M^2_R}$ and $M_2 = {\bf M^2_I}$, and substitute them into Eq.(15), we obtain several extra equations by
\begin{eqnarray}
{\bf A_1} &=& {\bf A_3 +B_2 (B_1^2 -B_3^2 )/B_1 B_3}, \nonumber \\
{\bf A_2} &=& {\bf A_3 +B_3 (B_1^2 -B_2^2 )/B_1 B_2}, \nonumber \\
{\bf C_2} &=& {\bf -B_3 C_3/B_2~~{\rm or}~~~x \equiv B_2 /B_3 =-C_3/C_2}, \nonumber \\
{\bf C_1} &=& ~~~{\bf B_3 C_3 /B_1 ~~{\rm or}~~~y \equiv B_1 /B_3 = ~~ C_3/C_1},
\end{eqnarray}
which reduce the number of independent parameters in $\bf M^2$ further down to five, and analytical diagonalization of $\bf M^2$ now becomes possible.
These equations look exactly the same as those previously given by Eq.(11)-(14) in \cite{Lin2019};
however, those were derived from an assumed Hermitian $M^q$ while these are derived from a naturally Hermitian $\bf M^2$.  \\

With the help of Eq.(16), ${\bf M^2}$ is now modified to
\begin{eqnarray}
{\bf M^2} =
\left( \begin{array}{ccc} {\bf A + x B (y- {1 \over y})}  & {\bf y B}    & {\bf x B}   \\  {\bf y B}  & {\bf A +B ({y \over x}-{x \over y})}  &  {\bf B} \\ {\bf x B}  & {\bf B}                & {\bf A}   \end{array}\right)
+ \left( \begin{array}{ccc} 0     & i {\bf C \over y}     & -i {\bf C \over x}   \\   -i {\bf C \over y}      & 0  &  i {\bf C} \\i {\bf C \over x}   & -i {\bf C}  & 0   \end{array}\right), \nonumber \\
\end{eqnarray}
if we choose five parameters ${\bf A \equiv A_3}$, ${\bf B \equiv B_3}$, ${\bf C \equiv C_3}$, ${\bf x \equiv B_2 / B_3}$ and ${\bf y \equiv  B_1 / B_3}$ to remain independent. \\

Then, eigenvalues of Eq.(17) are given by
\begin{eqnarray}
{\bf m^2_1} &=& {\bf A-B {x \over y} -C {\sqrt{\bf x^2 +y^2 +x^2 y^2} \over {x y}}}, \nonumber \\
{\bf m^2_2} &=& {\bf A-B{x \over y} + C{\sqrt{\bf x^2 +y^2 +x^2 y^2} \over {x y}}},  \nonumber \\
{\bf m^2_3} &=& {\bf A+B{{(x^2+1) y} \over x}},
\end{eqnarray}
with an $\bf U^q$ matrix given by
\begin{eqnarray}
{\bf U^q} = \left( \begin{array}{ccc}
{-\sqrt{\bf x^2+y^2} \over \sqrt{\bf 2(x^2+y^2+x^2 y^2)}} & {\bf {x(y^2-i \sqrt{\bf x^2+y^2+x^2 y^2})} \over {\bf \sqrt{2} \sqrt{\bf x^2+y^2} \sqrt{\bf x^2+y^2+x^2 y^2}}} & {\bf {y(x^2+i \sqrt{\bf x^2+y^2+x^2 y^2})} \over {\bf \sqrt{\bf 2} \sqrt{\bf x^2+y^2} \sqrt{\bf x^2+y^2+x^2 y^2}}}  \\
 {-\sqrt{\bf x^2+y^2} \over \sqrt{\bf 2(x^2+y^2+x^2 y^2)}} & {{\bf x(y^2+i \sqrt{\bf x^2+y^2+x^2 y^2})} \over {\sqrt{\bf 2} \sqrt{\bf x^2+y^2} \sqrt{\bf x^2+y^2+x^2 y^2}}} &~{{\bf y(x^2-i \sqrt{\bf x^2+y^2+x^2 y^2})} \over {\sqrt{\bf 2} \sqrt{\bf x^2+y^2} \sqrt{\bf x^2+y^2+x^2 y^2}}} \\
 {{\bf x y} \over \sqrt{\bf x^2+y^2+x^2 y^2}} &~ {\bf y \over \sqrt{\bf x^2+y^2+x^2 y^2}} &~{\bf x \over \sqrt{\bf x^2+y^2+x^2 y^2}} \end{array}\right).  \nonumber \\
\end{eqnarray}

It's astonishing that Eq.(19) depends on only two of those five remaining parameters.
As to be shown in next section, such a compact $\bf U^q$ pattern will give us a CKM matrix that depends on only four parameters.
Such parameterization is very natural since its elements are fully expressed in elements of $M^u$ and $M^d$. \\

In \cite{Lin1988, Lin2020}, four similar but simpler $U^q$ matrices have been obtained in $S_3$- and $S_2$-symmetric models.
They satisfy the necessary but not sufficient conditions stated in \cite{Lin2019} for yielding a complex CKM matrix.
They are in fact special cases of what obtained here with specific parameter values.
For instance, the $S_3$-symmetric case given in \cite{Lin1988} corresponds to the values $x=y=1$;
the $S_2$-symmetric cases given in \cite{Lin2020} correspond to values $x=-y=1$ (case 2), $x=-y=-1$ (case 3) and $x=y=-1$ (case 4 but with B and C replaced by their opposites), respectively. \\

In this section we replace the previously assumed Hermitian $M^q$ by a naturally Hermitian $\bf M^2$ to increase the generality of our investigation.
The only assumption employed here is the existence of a $\bf U^q$ matrix which diagonalizes the real part and imaginary part of ${\bf M^2}$ simultaneously and respectively.
With such an assumption, Eq.(15) brings in extra equations to simplify ${\bf M^2}$ down to that given in Eq.(17).
This enables us to diagonalize Eq.(17) analytically.
However, such a common $\bf U^q$ does not always exist.
For example, it does not exist in the two-dimensional case, of a $2 \times 2$ complex $\bf M^2$, unless either ${\bf M^2_R}$ or ${\bf M^2_I}$ is zero.
However, in the three-dimensional case, such a $\bf U^q$ has been proved to exist definitely as shown above.  \\

%%%%%%%%%%%%%%%%%%%%%
%%%  subsec: HTM  %%%
%%%%%%%%%%%%%%%%%%%%%
\section{The pattern of $\bf V_{CKM}$}

As the $\bf U^q$ pattern is obtained, it's natural to further study the $V_{CKM}$ pattern which is a product of two such matrices.
If we let the up-type $\bf U^u$ have the pattern given in Eq.(19) and the down-type $\bf U^d$ have the same pattern but with their parameters replaced by primed ones like $\bf A'$, $\bf B'$, $\bf C'$, ${\bf x'}$ and ${\bf y'}$, respectively, a $V_{CKM}$ will be thus obtained by substituting them into Eq.(3). \\

\begin{widetext}
\begin{table}[tbp]
\centering
\begin{tabular}{|l|llllll|}
\hline  $u\setminus d$      & ~~~~~~~~~(123) & ~~~~~~~~~(231) & ~~~~~~~~~(312) & ~~~~~~~~~(213) & ~~~~~~~~~(132) & ~~~~~~~~~(321) \\
\hline
$\left( \begin{array}{ccc}  1 \\ 2 \\ 3 \end{array}\right) $   &
 $\left( \begin{array}{ccc}   r^*  & s    & p^* \\  s^*  & r    & p   \\ p'   & p'^* & q \end{array}\right)$  &
 $\left( \begin{array}{ccc}   s    & p^*  & r^* \\  r    & p    & s^* \\ p'^* & q    & p' \end{array}\right)$   &
 $\left( \begin{array}{ccc}   p^*  & r^*  & s   \\  p    & s^*  & r  \\  q    & p'   & p'^*  \end{array}\right)$   &
 $\left( \begin{array}{ccc}   s    & r^*  & p^* \\  r    & s^*  &  p   \\ p'^* & p'   &  q \end{array}\right)$  &
 $\left( \begin{array}{ccc}   r^*  & p^* & s     \\  s^*  & p   & r   \\ p'   & q & p'^* \end{array}\right)$ &
 $\left( \begin{array}{ccc}   p^*  & s    & r^* \\  p   & r    & s^*  \\ q   & p'^* & p' \end{array}\right)$   \\
$\left( \begin{array}{ccc}  2 \\ 3 \\ 1 \end{array}\right) $     &
 $\left( \begin{array}{ccc}   s^*  & r    & p \\  p'  & p'^*    & q   \\ r^*   & s & p^* \end{array}\right)$ &
 $\left( \begin{array}{ccc}   r    & p  & s^* \\  p'^*    & q    & p' \\ s & p^*    & r^* \end{array}\right)$ &
 $\left( \begin{array}{ccc}   p  & s^*  & r   \\  q    & p'  & p'^*  \\  p^*    & r^*   & s  \end{array}\right)$ &
 $\left( \begin{array}{ccc}   r    & s^*  & p \\  p'^*    & p'  &  q   \\ s & r^*   &  p^* \end{array}\right)$ &
 $\left( \begin{array}{ccc}   s^*  & p & r     \\  p'  & q   & p'^*   \\ r^*   & p^* & s \end{array}\right) $ &
 $\left( \begin{array}{ccc}   p  & r    & s^* \\  q   & p'^*    & p'  \\ p^*   & s & r^* \end{array}\right)$   \\
$\left( \begin{array}{ccc}  3 \\ 1 \\ 2 \end{array}\right) $
 &  $\left( \begin{array}{ccc}   p'  & p'^*  & q \\  r^*  & s    & p^*   \\ s^*   & r & p \end{array}\right)$ &
 $ \left( \begin{array}{ccc}   p'^*    & q  & p' \\  s    & p^*    & r^* \\ r & p    & s^* \end{array}\right)$ &
 $ \left( \begin{array}{ccc}   q  & p'  & p'^*   \\  p^*    & r^*  & s  \\  p    & s^*   & r  \end{array}\right)$ &
 $\left( \begin{array}{ccc}   p'^*  & p'  & q \\  s    & r^*  &  p^*   \\ r & s^*   &  p \end{array}\right)$ &
 $\left( \begin{array}{ccc}   p'  & q & p'^*     \\ r^*  & p^*   & s   \\ s^*   & p & r \end{array}\right)$ &
 $ \left( \begin{array}{ccc}   q  & p'^*   & p' \\  p^*  & s    & r^*  \\ p   & r & s^* \end{array}\right)$   \\
$\left( \begin{array}{ccc}  2 \\ 1 \\ 3 \end{array}\right) $     &
 $\left( \begin{array}{ccc}  s^*  & r  & p \\  r^*  & s    & p^*   \\ p'   & p'^* & q \end{array}\right)$ &
 $\left( \begin{array}{ccc}   r    & p   & s^* \\  s    & p^*    & r^* \\ p'^* & q    & p' \end{array}\right)$ &
 $\left( \begin{array}{ccc}   p  & s^*  & r   \\  p^*    & r^*  & s  \\  q    & p'   & p'^*  \end{array}\right)$ &
 $\left( \begin{array}{ccc}   r  & s^*  & p \\  s    & r^*  &  p^*   \\ p'^* & p'   &  q \end{array}\right)$ &
 $\left( \begin{array}{ccc}   s^* & p & r     \\ r^*  & p^*   & s   \\ p'   & q & p'^* \end{array}\right)$ &
 $\left( \begin{array}{ccc}   p  & r   & s^* \\  p^*  & s    & r^*  \\ q   & p'^* & p' \end{array}\right)$   \\
$\left( \begin{array}{ccc}  1 \\ 3 \\ 2 \end{array}\right) $      &
 $\left( \begin{array}{ccc}  r^*  & s  & p^* \\  p'  & p'^*    & q   \\ s^*   & r & p \end{array}\right)$      &
 $\left( \begin{array}{ccc}   s    & p^*   & r^* \\  p'^*    & q    & p' \\ r & p    & s^* \end{array}\right)$      &
 $ \left( \begin{array}{ccc}   p^*  & r^*  & s   \\  q   & p'  & p'^*  \\  p    & s^*   & r  \end{array}\right)$      &
 $\left( \begin{array}{ccc}   s  & r^*  & p^* \\  p'^*  & p'  &  q   \\ r & s^*   &  p \end{array}\right)$      &
 $ \left( \begin{array}{ccc}   r^* & p^* & s     \\ p'  & q   & p'^*  \\ s^*  & p & r \end{array}\right)$      &
  $ \left( \begin{array}{ccc}   p^*  & s   & r^* \\  q  & p'^*   & p'  \\ p   & r & s^* \end{array}\right)$   \\
$\left( \begin{array}{ccc}  3 \\ 2 \\ 1 \end{array}\right) $    &
 $\left( \begin{array}{ccc}  p'  & p'^*  & q \\  s^*  & r    & p   \\ r^*   & s & p^* \end{array}\right)$      &
 $\left( \begin{array}{ccc}   p'^*  & q   & p' \\  r    & p    & s^* \\ s & p^*  & r^* \end{array}\right)$      &
 $ \left( \begin{array}{ccc}   q  & p'  & p'^*   \\  p^*    & s^*  & r  \\  p^*    & r^*   & s  \end{array}\right)$      &
 $ \left( \begin{array}{ccc}   p'^* & p'  & q \\  r    & s^*  &  p   \\ s & r^*   &  p^* \end{array}\right)$      &
 $ \left  ( \begin{array}{ccc}   p' & q & p'^*     \\ s^*  & p   & r   \\ r^*   & p^* & s \end{array}\right)$      &
 $\left( \begin{array}{ccc}   q  & p'^*  & p' \\  p  & r    & s^*  \\ p^*   & s & r^* \end{array}\right)$   \\
\hline
\end{tabular}
\caption{\label{tab:i} Thirty-six candidate patterns of $V_{CKM}$.
In the first row, there are six designations of the up-type $\bf m^2_1$, $\bf m^2_2$, and $\bf m^2_3$ to physical quarks $m^2_u$, $m^2_c$, and $m^2_t$ from light to heavy.
In the first column, there are also six such designations for down-type quarks.}
\end{table}
\end{widetext}

However, there is the problem of which $m^2_i$ corresponds to which $m^2_q$.
There are six ways to find the corresponding items in the up-quark sector, and there are also six in the down-quark sector.
Thus, totally thirty-six candidate $V_{CKM}$ patterns are exhibited in Table I.
The full expressions of elements in them are given by
\begin{widetext}
\begin{eqnarray}
 r~ &=& {{\bf (x^2+y^2)(x'^2+y'^2)+(x x' +y y')(x y x' y' +\sqrt{\bf x^2+y^2+x^2 y^2} \sqrt{\bf x'^2+ y'^2 +x'^2 y'^2})}
      \over {2 \sqrt{\bf  x^2+y^2} \sqrt{\bf  x'^2+y'^2}\sqrt{\bf x^2+y^2+x^2 y^2} \sqrt{\bf x'^2+y'^2+x'^2 y'^2} }} \nonumber \\
   &+& i~\bf  {{ (x y' -x' y)(x' y' \sqrt{\bf x^2+y^2+x^2 y^2} +x y \sqrt{\bf x'^2 +y'^2 +x'^2 y'^2})}
      \over {2 \sqrt{\bf  x^2+y^2} \sqrt{\bf  x'^2+y'^2}\sqrt{\bf x^2+y^2+x^2 y^2}\sqrt{\bf x'^2+y'^2+x'^2 y'^2}}}, \\
 s~ &=& \bf {{(x^2+y^2)(x'^2+y'^2)+(x x' +y y')(x y x' y' -\sqrt{\bf x^2+y^2+x^2 y^2} \sqrt{\bf x'^2+ y'^2 +x'^2 y'^2})}
      \over {2 \sqrt{\bf  x^2+y^2} \sqrt{\bf  x'^2+y'^2}\sqrt{\bf x^2+y^2+x^2 y^2} \sqrt{\bf x'^2+y'^2+x'^2 y'^2} }} \nonumber \\
   &+& i~ \bf {{ (x y' -x' y)(x' y' \sqrt{\bf x^2+y^2+x^2 y^2} -x y \sqrt{\bf x'^2 +y'^2 +x'^2 y'^2})}
      \over {2 \sqrt{\bf  x^2+y^2} \sqrt{\bf  x'^2+y'^2}\sqrt{\bf x^2+y^2+x^2 y^2}\sqrt{\bf x'^2+y'^2+x'^2 y'^2}}}, \\
 p~ &=& \bf {{[y' y^2(x-x')+ x' x^2 (y-y')]+ i(x y'-x' y) \sqrt{\bf x^2+y^2+x^2 y^2}} \over {\sqrt{2} \sqrt{\bf x^2+y^2} \sqrt{\bf x^2+y^2+x^2 y^2}\sqrt{\bf x'^2+y'^2+x'^2 y'^2}}}, \\
 p' &=& \bf {{[y y'^2 (x'-x)+ x x'^2 (y'-y)]+ i(x y'-x' y) \sqrt{\bf x'^2+y'^2+x'^2 y'^2}} \over {\sqrt{2} \sqrt{\bf x^2+y^2+x^2 y^2} \sqrt{\bf x'^2+y'^2}\sqrt{\bf x'^2+y'^2+x'^2 y'^2}} }, \\
 q~ &=& \bf  {{x x'+y y' +x y x' y'}\over {\sqrt{\bf  x^2+y^2+x^2 y^2} \sqrt{\bf x'^2+y'^2+x'^2 y'^2}}},
\end{eqnarray}
\end{widetext}
in which all elements depend on only four parameters, and they are allowed to be complex if the parameters are properly chosen.
Such parameterization of $V_{CKM}$ is very natural since all of the elements are composed of the Yukawa couplings presented in Eq.(1) and the VEV of its only Higgs doublet.  \\

However, there is a very serious problem in these candidate CKM matrices.
If $V_{CKM}$ is unitary, the elements must obey following rules
\begin{eqnarray}
 2 \vert p \vert^2 + \vert q \vert^2 &=& 1, \\
 2 \vert p' \vert^2 + \vert q \vert^2 &=& 1, \\
  \vert p \vert^2 + \vert r \vert^2 + \vert s \vert^2  &=& 1,
\end{eqnarray}
and they give a quadruple equality
\begin{eqnarray}
   \vert p \vert = \vert p' \vert ,
\end{eqnarray}
since each of them correlates to two CKM elements.
Such a quadruple equality is a big problem since it must include CKM elements which are very different. \\

As we have get candidates which are shown in Table I, the next step is naturally to find out which of them fits experiments best.
The first reference standard used here is the empirical CKM elements given in \cite{Zyla2020}
\begin{eqnarray}
V_{CKM}^{emp.} &=& \left( \begin{array}{ccc}     0.97401^{+0.00011}_{-0.00011}   & 0.22650^{+0.00048}_{-0.00048} & 0.00361^{+0.00011}_{-0.00009} \\
                                   0.22636^{+0.00048}_{-0.00048} & 0.97320^{+0.00011}_{-0.00011} & 0.04053^{+0.00083}_{-0.00061}    \\
                                   0.00854^{+0.00023}_{-0.00016} & 0.03978^{+0.00082}_{-0.00060} & 0.999172^{+0.0000024}_{-0.000035}         \end{array}\right) \nonumber \\
     &\approx & \left( \begin{array}{ccc}     O(1)   & O(\lambda) & O(\lambda^3) \\
                                   O(\lambda) & O(1) & O(\lambda^2)    \\
                                   O(\lambda^3) & O(\lambda^2) & O(1)         \end{array}\right),
\end{eqnarray}
where $\lambda \approx 0.22$ is one of the Wolfenstein's parameters. \\

In Eq.(29), the values of $V_{CKM}$ elements can be classified into four grades $O(1),~O(\lambda),~ O(\lambda^2)$ and $O(\lambda^3)$.
If we classify the ratio of two elements on both sides of a predicted equality in the same way,
we may estimate the rationalities of each $V_{CKM}$ by the ratio of its most significant pair and the results are demonstrated in Table II.
For instance, in the $\left( \begin{array}{ccc}  1 \\ 2 \\ 3 \end{array}\right) ~(1~2~3)$ case, the ratio of $\vert V_{td} \vert$ to $\vert V_{ts} \vert$ is about $0.2147 \approx O(\lambda) $.
Among all thirty-six candidates, twenty-eight of them have at least one such pair whose ratio is of the order of $O(\lambda^2)$ or $O(\lambda^3)$.
Thus, they will be excluded in subsequent discussions, and only those eight of the order of $O(\lambda)$ will be considered. \\

\begin{table}[tbp]
\centering
\begin{tabular}{|l|llllll|}
\hline  $u\setminus d$      & ~~(123) & ~~(231) & ~~(312) & ~~(213) & ~~(132) & ~~(321) \\
\hline
$\left( \begin{array}{ccc}  1 \\ 2 \\ 3 \end{array}\right) $    &
 $\bf O(\lambda )$   & $O(\lambda^3 )$  &  $O(\lambda^3 )$   &  $\bf O(\lambda )$   &  $O(\lambda^3 )$   &  $O(\lambda^3 )$     \\
$\left( \begin{array}{ccc}  2 \\ 3 \\ 1 \end{array}\right) $    &
 $O(\lambda^3 )$   & $\bf O(\lambda )$  &  $O(\lambda^3 )$   &  $O(\lambda^3)$   &  $\bf O(\lambda)$   &  $O(\lambda^3 )$  \\
$\left( \begin{array}{ccc}  3 \\ 1 \\ 2 \end{array}\right) $    &
 $O(\lambda^3 )$   & $O(\lambda^3 )$  &  $O(\lambda^2 )$   &  $O(\lambda^3)$   &  $O(\lambda^3)$   &  $O(\lambda^2 )$   \\
$\left( \begin{array}{ccc}  2 \\ 1 \\ 3 \end{array}\right) $    &
 $\bf O(\lambda )$   & $O(\lambda^3 )$  &  $O(\lambda^3 )$   &  $\bf O(\lambda)$   &  $O(\lambda^3)$   &  $O(\lambda^3 )$   \\
$\left( \begin{array}{ccc}  1 \\ 3 \\ 2 \end{array}\right) $    &
 $O(\lambda^3 )$   & $\bf O(\lambda )$  &  $O(\lambda^3 )$   &  $O(\lambda^3)$   &  $\bf O(\lambda)$   &  $O(\lambda^3 ) $  \\
$\left( \begin{array}{ccc}  3 \\ 2 \\ 1 \end{array}\right) $    &
 $O(\lambda^3 )$   & $O(\lambda^3 )$  &  $O(\lambda^2 )$   &  $O(\lambda^3)$   &  $O(\lambda^3)$   &  $O(\lambda^2 )$  \\
\hline
\end{tabular}
\caption{\label{tab:i} Estimations of irrationalities of equalities between CKM elements.
If we classify the $V_{CKM}$ elements into several grades like $V_{ud} \approx V_{cs} \approx V_{tb} \approx O(1)$, $V_{us} \approx V_{cd} \approx O(\lambda)$, $V_{cb} \approx V_{ts} \approx O(\lambda^2)$, and $V_{ub} \approx V_{td} \approx O(\lambda^3)$, the notations given above denote the largest differences between the predicted pairs in order of Wolfenstein's $\lambda$ parameter.     }
\end{table}

However, these eight $O(\lambda)$-level $V_{CKM}$ are further found to be grouped into four complex conjugate pairs as shown below
\begin{eqnarray}
V{\left( \begin{array}{ccc}  2 \\ 1 \\ 3 \end{array}\right) ~(1~2~3)} &=& V^*{\left( \begin{array}{ccc}  1 \\ 2 \\ 3 \end{array}\right) ~(2~1~3)}
 =  \left( \begin{array}{ccc}  s^*  & r  & p \\  r^*  & s    & p^*   \\ p'   & p'^* & q \end{array}\right), \\
V{\left( \begin{array}{ccc}  1 \\ 2 \\ 3 \end{array}\right) ~(1~2~3)} &=& V^*{\left( \begin{array}{ccc}  2 \\ 1 \\ 3 \end{array}\right) ~(2~1~3)}
 =\left( \begin{array}{ccc}   r^*  & s    & p^* \\  s^*  & r    & p \\ p'  & p'^* & q \end{array}\right), \\
V{\left( \begin{array}{ccc}  1 \\ 3 \\ 2 \end{array}\right) ~(2~3~1)} &=& V^*{\left( \begin{array}{ccc}  2 \\ 3 \\ 1 \end{array}\right) ~(1~3~2)}
 =\left( \begin{array}{ccc}   s    & p^*   & r^* \\  p'^*    & q    & p' \\ r & p    & s^* \end{array}\right), \\
V{\left( \begin{array}{ccc}  2 \\ 3 \\ 1 \end{array}\right) ~(2~3~1)} &=& V^*{\left( \begin{array}{ccc}  1 \\ 3 \\ 2 \end{array}\right) ~(1~3~2)}
 =\left( \begin{array}{ccc} r  & p  & s^* \\  p'^*    & q    & p' \\ s & p^*    & r^* \end{array}\right).
\end{eqnarray}

Besides, these four pairs can be even further grouped into two sets in which one of the members can be obtained from the other by exchanging their $r$ and $s$.
For instance, exchanging the $r$ and $s$ parameters in Eq.(30) and (32) will give Eq.(31) and (33), respectively.
Thus, those eight remaining candidates can be further classified into two different groups. \\

As we get these two groups of $V_{CKM}$ candidates, we perform numerical tests to find out in what kind of parameter spaces they will fit empirical values best.
In those cases given in Eq.(30) and (31), we can not find a parameter space in which $\vert r \vert \geq 0.89$ and $0.3 \geq \vert s \vert \geq 0.1$ while $\vert V_{cb} \vert (\approx 0.04053) \geq \vert p \vert \geq \vert V_{ub} \vert (\approx 0.00365)$, or $\vert s \vert \geq 0.89$ and $0.3 \geq \vert r \vert \geq 0.1$ while $\vert p \vert$ falls in the same range.
While in cases of Eq.(32) and (33), there are many sets of ${\bf x}$, ${\bf y}$, ${\bf x'}$ and ${\bf y'}$ which all give the same set of results
\begin{eqnarray}
\vert s \vert &~& ({\rm or} \vert r \vert) = \vert V_{ud} \vert = \vert V_{tb} \vert \approx ~0.9925, \\
\vert r \vert &~& ({\rm or} \vert s \vert) = \vert V_{ub} \vert = \vert V_{td} \vert \approx ~0.0075, \\
\vert p \vert &~& = \vert p' \vert = \vert V_{us} \vert = \vert V_{ts} \vert = \vert V_{cd} \vert = \vert V_{cb} \vert \approx  0.122023, \\
\vert q \vert &~& = \vert V_{cs} \vert \approx  0.9845,
\end{eqnarray}
in the range from -25 to +25 of each parameter.  \\

Taking Eq.(32) as an example and compare those values in Eq.(34)-(37) with Eq.(29), we find that
$\vert s \vert =0.9925$ in Eq.(34) is very close to $\vert V_{ud} \vert^{emp.} = 0.97401$ and $ \vert V_{tb} \vert^{emp.} = 0.999172$,
$\vert r \vert =0.0075$ in Eq.(35) is very close to $\vert V_{ub} \vert^{emp.} = 0.00361$ and $ \vert V_{td} \vert^{emp.} = 0.00854$, and
$\vert q \vert =0.9845$ in Eq.(37) is very close to $\vert V_{cs} \vert^{emp.} = 0.97320$.
These predictions are already very close to the empirical values. \\

On the other hand,
the quadruple equality predicted in Eq.(36) has the same value 0.122023 which lies between the $O(\lambda)$ pair $\vert V_{us} \vert ~\approx ~\vert V_{cd} \vert ~\approx 0.226$
and the $O(\lambda^2)$ pair $\vert V_{ts} \vert ~\approx ~ \vert V_{cb} \vert ~\approx 0.04$.
Though it's still a little far from both ends,
however, it's very close to the intermediate value 0.13314 of the largest $\vert V_{us} \vert = 0.2265$ and the smallest $\vert V_{ts} \vert =0.03978$ of the quadruplets,
or 0.09492 the geometric mean of them.
Thus, we will study this set of $V_{CKM}$ candidates more in what follows. \\

It's very interesting that in Eq.(32) and (33) the designations of eigenvalues to physical quarks in up- and down-type sectors are different.
For instance, the $V{\left( \begin{array}{ccc}  1 \\ 3 \\ 2 \end{array}\right) ~(2~3~1)}$ case in Eq.(32) corresponds to
\begin{eqnarray}
(m^2_u,~m^2_c,m^2_t) &\leftrightarrow & ~({\bf m^2_1,~m^2_3,~m^2_2}) \nonumber \\
(m^2_d,~m^2_s,m^2_b) &\leftrightarrow & ~({\bf m'^2_2,~m'^2_3,~m'^2_1}),
\end{eqnarray}
while the $V^*{\left( \begin{array}{ccc}  2 \\ 3 \\ 1 \end{array}\right) ~(1~3~2)}$ case corresponds to
\begin{eqnarray}
(m^2_u,~m^2_c,m^2_t) &\leftrightarrow & ~({\bf m^2_2,~m^2_3,~m^2_1}) \nonumber \\
(m^2_d,~m^2_s,m^2_b) &\leftrightarrow & ~({\bf m'^2_1,~m'^2_3,~m'^2_2}).
\end{eqnarray}

In Eq.(33), the $V{\left( \begin{array}{ccc}  2 \\ 3 \\ 1 \end{array}\right) ~(2~3~1)}$ case corresponds to
\begin{eqnarray}
(m^2_u,~m^2_c,m^2_t) &\leftrightarrow & ~({\bf m^2_2,~m^2_3,~m^2_1}) \nonumber \\
(m^2_d,~m^2_s,m^2_b) &\leftrightarrow & ~({\bf m'^2_2,~m'^2_3,~m'^2_1}).
\end{eqnarray}
while the $V^*{\left( \begin{array}{ccc}  1 \\ 3 \\ 2 \end{array}\right) ~(1~3~2)}$ case corresponds to
\begin{eqnarray}
(m^2_u,~m^2_c,m^2_t) &\leftrightarrow & ~({\bf m^2_1,~m^2_3,~m^2_2}) \nonumber \\
(m^2_d,~m^2_s,m^2_b) &\leftrightarrow & ~({\bf m'^2_1,~m'^2_3,~m'^2_2}).
\end{eqnarray}

It's interesting that there is a commonality among Eq.(34)-(37);
the eigenvalues $\bf m^2_3$ and $\bf m'^2_3$ always correspond to the middle ones, $m^2_c$ and $m^2_s$, respectively, in both quark types.
The differences among those four equations lies only in taking up-type quarks for an example, how $\bf m^2_1$ and $\bf m^2_2$ should be, respectively, assigned to $m^2_u$ and $m^2_t$,
and the same for the down-quark sector.
However, as shown in Eq.(18), the $(m^2_u,~m^2_t) \leftrightarrow ({\bf m^2_1,~ m^2_2})$ case can be obtained from the $(m^2_u,~m^2_t) \leftrightarrow ({\bf m^2_2,~ m^2_1}) $ case by giving the parameter $\bf C$ a "-" sign.  \\

The difference between $\bf m^2_1$ and $\bf m^2_2$ is just the $\bf C$ dependent term, $\bf \pm C {\sqrt{\bf x^2 +y^2 +x^2 y^2} \over {x y}}$.
It's more interesting that if $\bf C$=0, $m^2_u$ and $m^2_t$ will be equal, or degenerate.
We may imagine that $m^2_u$ and $m^2_t$ were degenerate when $\bf C$=0 in some very early stage of our universe.
They were split up into two different values when $\bf C$ became nonvanishing and grew to present values.
This suggests that quark masses could be not constants or they are $running$ in time.
Unfortunately, those equations do not tell us how they run with time $\bf t$ or temperature $\bf T$ of the universe.
\\

As mentioned above, there are many sets of parameters giving numerically the same set of $p$, $p'$, $q$, $r$, and $s$ values shown in Eq.(34)-(37) and we do not know which one is the best solution.
Here, we would like to pick up one of them,
\begin{eqnarray}
{\bf x} &=& -16.3201, ~{\bf y}=19.767, \nonumber \\
{\bf x'} &=& 12.4127,~{\bf y'}=-20.001,
\end{eqnarray}
for further studies on their values. \\

If we let ${\bf m^2_1}=m^2_u$, ${\bf m^2_2}=m^2_t$, ${\bf m'^2_1}=m^2_d$, and ${\bf m'^2_2}=m^2_b$ and substitute the empirical quark masses
$m_u =0.00216$ GeV, $m_c=1.27$ GeV, $m_t=172.76$ GeV, $m_d=0.00467$ GeV, $m_s=0.093$ GeV, and $m_b=4.18$ GeV into Eq.(18), we will get
\begin{eqnarray}
{\bf A} &=& 14885.1,~~~{\bf B}=45.9634,~~~~~~{\bf C}=-14876.1, \\
{\bf A'} &=& 8.71459,~~~{\bf B'}=0.0348408,~~~{\bf C'}=-8.69718,
\end{eqnarray}
which are their present values.
Even if these six parameters are fixed or constants, the quark masses vary with the other four $\bf x$, $\bf y$, $\bf x'$, and $\bf y'$ parameters.
This describes the $running$ of quark masses in another way with very interesting theoretical descriptions. \\

In this section, a very general CKM pattern is given analytically.
It's still not the ultimate solution since there is still one assumption employed.
From our experiences obtained in previous investigations, from $S_3$ symmetry \cite{Lin1988} to $S_2$ symmetry \cite{Lin2020} to "no symmetry" \cite{Lin2019},
the ${\bf M^2}$ pattern given in Eq.(17) is still oversimplified.
That is why there is a quadruple equality in the derived $V_{CKM}$.
We expect that directly diagonalization of Eq.(4) may solve this problem completely.
Unfortunately, such an ultimate solution looks still unattainable for now.
We still need more efforts on this subject.  \\

\section{Conclusions and Discussions}

The $V_{CKM}$ presented here is so general that it's already only one step away from the ultimate solution of CPV problem in the standard model,
since only one assumption is employed in this research.
The natural Hermitian $\bf M^2$ studied here has only nine independent parameters,
 and the assumption of a common $\bf U^q$ further reduce the parameter number down to five. \\

Analytical diagonalization of such a matrix gives three eigenvalues, among which two are degenerate at $\bf C$=0 and a $\bf U^q$ depends on only two of the parameters.
With such a $\bf U^q$ pattern, thirty-six $V_{CKM}$ candidates are yielded.
Twenty-eight of them are excluded in the first run of examination, and the remaining eight are divided into two groups.
One of the remaining two groups is further excluded by a numerical test.
\\

The finally remaining group of $V_{CKM}$ candidates predicts several equalities among $V_{CKM}$ elements.
One of them is a quadruple equality among four elements, which includes elements of very different values.
In a numerical test, only the finally remaining group gives unsatisfactory but acceptable values for CKM elements in that equality.
This suggests that  hints the $V_{CKM}$ thus derived is not the ultimate one and the assumption of a common $\bf U^q$ still oversimplifies the $\bf M^2$ pattern.
Besides, the parameterization of $V_{CKM}$ given here is a very natural one since all its elements are dependent on only four parameters which are dependent only on Yukawa couplings and VEV of the only Higgs doublet in SM. \\

However, the fitting of derived $V_{CKM}$ elements with empirical values indicates those two quarks predicted to be degenerate at $\bf C$=0 are the lightest and heaviest generations in a quark type, say $m_u =m_t$ in the up-quark sector, rather than the lighter two generations supposed in our previous researches.
Besides, the derived eigenvalues allow quark masses to be running in history of the universe, since under a completely $S_3$-symmetric circumstance the values $x=1$, $y=1$, $x'=1$ and $y'=1$ are very different to those given in Eq.(42).
This is a very interesting topic to be studied in future researches. \\

Though this research does not solve the CPV problem completely, it is already one step away from the ultimate solution of it.
The only barrier now on the way is the assumption of a common $\bf U^q$ which diagonalizes $\bf M^2_R$ and $\bf M^2_I$ simultaneously and respectively.
If we can diagonalize the unsimplified $\bf M^2$ directly, surely the problem will be solved completely.
Before such an ultimate solution comes onto the stage, the $V_{CKM}$ presented here is the closest one to it. \\

\end{document}